\documentstyle[amsfonts,preprint,aps]{revtex}
\begin{document}

\title{
Order versus Disorder in the Quantum Heisenberg Antiferromagnet
on the Kagom\'e lattice: an approach through exact spectra 
analysis.}
\author{
P. Lecheminant \thanks{Groupe
    de Physique Statistique, Universit\'e de Cergy-Pontoise,
    site de Saint-Martin, 2 avenue Adolphe Chauvin,
    95302 Cergy-Pontoise Cedex, France.},
B. Bernu \thanks{Laboratoire de Physique Th\'eorique des
    Liquides, Universit\'e P. et M. Curie, boite 121, 4 Place Jussieu, 75252
    Paris Cedex, France. URA 765 of CNRS},
C. Lhuillier $ ^{ \dag} $, and
L. Pierre\thanks{
     Universit\'e Paris-X, Nanterre, 92001 
Nanterre Cedex, France.},
P. Sindzingre  $ ^{ \dag} $\\
}
\date{\today}
\maketitle

\bibliographystyle{prsty}

\begin{abstract}
A group symmetry analysis of the low lying levels
of the spin-1/2 kagom\'e Heisenberg antiferromagnet 
is performed for small samples 
up to $N=27$. This new approach allows to follow 
the effect of quantum fluctuations when the 
sample size increases. The results contradict the 
scenario of ``order by disorder'' which has been 
advanced on the basis of large S calculations. 
A large enough second neighbor 
ferromagnetic exchange coupling is needed to stabilize
the $\sqrt 3 \times \sqrt 3$ pattern: the  finite size 
analysis indicates a quantum critical transition at a non zero coupling.
\end{abstract}

\newpage
\narrowtext

\section{Introduction}
        There are very few simple two dimensional magnets which fail to
order at $T=0$. There is now a large amount of evidence that the $S=1/2$
nearest neighbor Heisenberg antiferromagnet is ordered not only on the square
lattice\cite{m91} 
but also on the triangular lattice (THA) 
\cite{he88,jlg89,blp92,de93,esy93,bllp94,lblp195,bo95}.
The reduction of the order parameter by quantum
fluctuations is about $40\%$ for the square lattice 
and of the order of $50\%$
for the triangular lattice: frustration indeed enhances the role of the
quantum fluctuations but the relatively high coordination number plays
against them. The kagom\'e lattice which can be seen as a diluted triangular
lattice (see Fig. \ref{fig:res}) 
exhibits both frustration and low coordinance number
and the $S=1/2$ Heisenberg antiferromagnet on the kagom\'e (KAH) 
is a good candidate for a disordered two dimensional quantum
liquid. 

Exact diagonalizations on periodic samples have shown that the spin-spin
correlations decrease indeed very rapidly with distance 
\cite{ze90,ce92,le93}.
Series expansion from the Ising limit and high temperature series 
point to the absence of magnetic order\cite{sh92,el94}. Large-N
approaches, using Sp(N) bosons\cite{s92}, find a 
disordered  ground-state with unbroken symmetry for 
small enough spin while, using SU(N) fermions, a 
spin-Peierls phase or a chiral phase 
is suggested\cite{mz91}. The best variational
energies for the N=36 sample are built from 
the short-ranged dimerized basis\cite{ze95}. 
 
On the other hand, semiclassical 
approaches\cite{hkb92,c92,c193,mtgc94,as95}
plead in favor of a magnetic order 
(the $\sqrt 3\times \sqrt 3$ state, see Fig. \ref{fig:res})
induced by quantum fluctuations (``order
by disorder''\cite{vbcc80}). This kind of phenomenon 
has already been seen in much 
simpler situations where there is a classical degeneracy between two kinds
of orders\cite{s82,h89}. The $J_1-J_2$ model on the 
triangular lattice for large enough $J_2$ ($J_2/J_1 > 1/8$) possesses 
classically degenerate ordered  ground-states with respectively two and four
sublattices\cite{jdgb90}. Large spin calculations have predicted that
quantum fluctuations should select the two sublattices order\cite{cj92a,k93}.
The study of small samples spectra has confirmed
this prediction and shown the mechanism of this selection\cite{lblp295}. 

The kagom\'e antiferromagnet is a more problematic issue 
because of the infinite
number of classically degenerate  ground-states\cite{chs92,ccr93}.
The linear spin wave approach for the kagom\'e
antiferromagnet does not lift the
extensive degeneracy of the
classical  ground-state (at this order the spectrum of magnetic excitations
possesses a whole branch of zero modes\cite{hkb92,chs92,ccr93}).
One has to invoke non linear
processes to stabilize the $\sqrt 3\times \sqrt 3$
ordered solution\cite{c92,c193,as95}.
On the other hand, one should note that in this class 
of problem there exist some situations 
where the ``order by disorder'' phenomenon
fails for systems with a  ground-state manifold associated with 
an extensive entropy. An example is the quantum Heisenberg antiferromagnet
on the triangular Husimi cactus\cite{cd94} where the system prefers 
to remain a spin liquid rather than localizing in a particular 
ordered state that breaks the degeneracy of the classical 
ground state manifold. Moreover, the possibility of a quantum 
tunnelling between different classical states might prevent 
the system from localizing in a particular 
ordered solution\cite{dh92,dh93}. 

This paper is a scrutiny of the  conjecture of selection of order by quantum
fluctuations based on the study of the low lying levels of exact spectra.
The paper is organized as follows:
A brief presentation of our method to investigate the finite 
size properties of an ordered antiferromagnet is given in 
the next section. In Sec. III, we apply this method to the search 
of the selection
of the $\sqrt 3\times \sqrt 3$ state
by quantum fluctuations.
In Sec. IV, we introduce a second neighbor ferromagnetic 
exchange coupling ($J_2 < 0$) and show that for large enough
$|J_2|$, the system orders through the $\sqrt 3\times \sqrt 3$ state.   
The rigidity of the KHA and the hypothesis of an incommensurable
planar order are investigated in Sec. V. Section VI, finally, 
summarizes our results and lists some open questions. 

\section{Review of the method}
       The presence of order in a quantum antiferromagnet is readily seen by
examination of the low lying levels of its spectrum. The method used in
this work has been described in details in \cite{blp92,bllp94,lblp295},
we give here a summary of its most important features.
        The first characteristic of an ordered antiferromagnet (classical
or quantum) is the existence of ferromagnetic sublattices: the total spin
of each ferromagnetic sublattice is the collective variable relevant for
the description of the low energy spectrum of the system. On a finite
lattice with $p$ sublattices of $Q$ sites (the total number of sites 
being $N = pQ$),
the $p$ identical spins $Q/2$ couple to form the rotationally invariant states
that are the low lying eigenstates of the hamiltonian. These $p$ coupled
spins form a ``generalized top'' which  precesses freely: elementary mechanics
(confirmed by  the non-linear sigma model approach \cite{f89,nz89,adm93}) 
indicates that the leading term of this precession is:
\begin{equation}
H_{\rm eff}= \frac{{\bf S}^{2}}{2I_{N}}.
\label{eq-top1}
\end{equation}
In an ordered system,  $I_N$, the inertia of the top 
is an extensive quantity,
proportional to the perpendicular susceptiblity of the magnet 
\cite{blp92,adm93}. In a disordered system with a gap
this quantity is asymptotically constant, and at the quantum critical 
point between this two regimes $I_N$ is expected to vary as 
$N^{-1/2}$\cite{adm93}. The coupling of the $pQ/2$ spins gives
$N_S(N,S,p)$ states for each $S$
value of the total spin of the sample and a total of $(Q+1)^p$ levels obeying
the effective dynamics of Eq. (\ref{eq-top1}). 
These levels form the quantum counterpart
of the classical  ground-state: they  represent the tower of states indicated
by Anderson in his 1952 famous paper \cite{a52}. In our spectral
representation  where the eigenstate energies are displayed as a
function
of $S(S+1)$ in order to exhibit the free precession (Eq. (\ref{eq-top1})), 
the tower is in
fact a ``Pisa Tower'' with a slope decreasing with the sample size.
In the thermodynamic limit, all these low lying
levels (called QDJS  in Ref. \cite{blp92} for 
quasi-degenerate joint states)
collapse to the absolute  ground-state as $N^{-1}$.  An experimental proof of
these levels has recently been reported in the analysis of macroscopic
quantum coherence in antiferromagnets\cite{lh95}.
It could be noticed that in the thermodynamic limit these
levels give a   ground-state entropy proportional to  $\ln(N^p)$. 
Above these first family of levels, the spectrum of an ordered
antiferromagnet exhibits the magnon spectrum; the low lying levels of this
part of the spectrum collapse more slowly than the QDJS to the  ground-state
with a scaling law $N^{-1/2}$.
Three features of the ``Pisa Tower'' of QDJS are thus essential:

\noindent
$i)$ the overall effective dynamics of a finite family of levels (of the
order of $N^{p}$) and its finite size scaling leading 
to a clear cut separation
from the first inhomogeneous magnon excitations
(the absence of separation
between the scaling laws would sign a quantum critical 
behaviour \cite{adm93}),\\
$ii)$ the numbers of levels $N_S(N,S,p)$ in each 
$S$ subset which for given $N$ and $S$ is determined by  the number
$p$ of
ferromagnetic sublattices of the underlying N\'eel order,\\
$iii)$ the spatial symmetries of these $N_S(N,S,p)$ levels:
 The number and nature of the space group irreducible representations
(S.G.I.R.)
that appear in each S subspace are uniquely determined by the
geometrical symmetries of the N\'eel order.

The exact diagonalization of the Heisenberg hamiltonian on 
small samples enables us to examine these features and  to determine 
the nature of the ordering.
This approach has been used for the quantum Heisenberg antiferromagnet
on the triangular lattice\cite{blp92,bllp94} 
and the $J_1-J_2$ problem on the triangular
lattice\cite{lblp295}.
 
\section{The low lying levels of the KAH} 

The hamiltonian of the KAH, including further neighbor interactions,
has the form: 
\begin{equation}
{\cal H} = 2 J_1 \sum_{<i,j>} {\bf S}_i . {\bf S}_j +
2 J_2 \sum_{<<i,k>>} {\bf S}_i . {\bf S}_k
\label{eq-ham}
\end{equation}
where the ${\bf S}_i$ are spin-1/2 quantum operators on the
sites of the kagom\'e lattice and $<i,j>$ (resp. $<<i,k>>$)
denotes a sum over first (resp. second) neighbors. In this 
section, we shall only consider first neighbor exchange interactions
($J_2 = 0$). 
At  first sight, the spectrum of the KAH has one feature which is
common to all the Heisenberg  antiferromagnet 
spectra that we have been studying: the
absolute  ground-state has a total spin $S=0$ or $1/2$ depending on the
number of sites of the sample ($N=9,12,15,18,21,24,27$) 
and the   ground-state energy $E_0(S)$ 
of the $S$ sectors order with increasing $S$ values.
Lieb and Mattis\cite{lm62} have shown that this result is exact for
bipartite lattices: our numerical results tend to indicate an
extremely robust property (the theorem seems to be true for the Heisenberg
antiferromagnet on the triangular and  kagom\'e lattices and for 
the $J_1-J_2$ 
problem: none of these situations can be reduced to a bipartite problem).
Taken apart this feature the spectra of the KAH appear totally
different from the spectra of the TAH. In Fig. \ref{fig:KAH}, the low lying
levels of the TAH and of the KAH 
are shown for the $27$ sites samples: 

\noindent
$i)$ whereas the ``Pisa Tower'' is easily seen on the TAH spectrum, 
well separated from
the magnons spectrum, there is absolutely no such scales in the KAH
one,\\
$ii)$ the effective dynamics of the low lying levels of the KAH spectrum do
not scale as $S(S+1)$,\\ 
$iii)$ the symmetries of the lowest lying levels of each $S$ subspace 
do not allow the description of an ordered structure: for $N=27$,  
all the S.G.I.R.,  but one, appear in the
low lying doublet states below the first $S=3/2$ eigenstates, whereas
their
number $N_S$ and nature are strictly determined in the case of an
ordered solution.\\
One could argue that the proliferation of these low lying levels
are the quantum counterpart of the infinite degeneracy of the classical
ground state with respect to local spin rotations.
The real question is: do the quantum fluctuations show any trend to
select a specific N\'eel order? 

We have looked to this question
for the so-called ${\bf q} = {\bf 0}$ 
order,  the $\sqrt 3 \times \sqrt 3$ order
and for any planar order (see section V).
The ${\bf q} = {\bf 0}$  order is studied in  Ref. \cite{phle95}.
We give here
the details of the analysis concerning 
the $\sqrt 3 \times \sqrt 3$ order, which is
the favored solution found 
in the semi classical approaches\cite{hkb92,c92,c193,mtgc94,as95}.
The smallest lattices where periodic boundary conditions are
compatible with this order are $N=9,27$ and $36$ sites. 
In this section, we shall consider explicitly the $N=27$ 
sample since the $N=9$ sites is too small and the $N=36$ sample
is too large to compute all the levels in each $S$ sector.
The QDJS associated with the $\sqrt 3 \times \sqrt 3$ state 
are homogeneous on each ferromagnetic sublattices (their
wave-vectors are either ${\bf k} = {\bf 0}$ 
or ${\bf k} =\pm {\bf k_0}$: corners of the Brillouin zone). They
do not break the $C_{3v}$ symmetry of the lattice.
The three irreducible representations
characterizing the $\sqrt 3 \times \sqrt 3$ order are the following:
\begin{equation}
\left\{
\begin{array}{l}
\displaystyle \Gamma_1 : \left[{\bf k} = {\bf 0}, {\cal R}_\pi\Psi=\Psi,
{\cal R}_{2\pi/3}\Psi=\Psi, \sigma_x\Psi=\Psi \right]
\\
\displaystyle \Gamma_2 : \left[{\bf k} = {\bf 0}, {\cal R}_\pi\Psi=-\Psi,
{\cal R}_{2\pi/3}\Psi=\Psi, \sigma_x\Psi=\Psi \right]
\\
\displaystyle \Gamma_3 : \left[{\bf k} = \pm {\bf k}_{0},
{\cal R}_{2\pi/3}\Psi=\Psi, \sigma_x\Psi=\Psi \right],
\end{array}
\right.
\label{eq-gamari}
\end{equation}
where ${\cal R}_{\phi}$ is a rotation of angle $\phi$ 
and $\sigma_x$ denotes an axial symmetry.
The numbers $N_S\left(N,S,p=3\right)$  of levels in the ``Pisa Tower''
for  each value of the total spin are given
by the coupling of three $N/6$ spins: 
\begin{equation}  
{\cal D}^{N/6} \otimes {\cal D}^{N/6} \otimes {\cal D}^{N/6} = 
\sum_{S=0}^{N/6} \left(2S+1\right) {\cal D}^S +
\sum_{S=N/6+1}^{N/2} \left(N/2-S+1\right) {\cal D}^S
\end{equation}  
where ${\cal D}^S$ denotes the irreducible representation for
a spin $S$. Therefore, one obtains the numbers $N_S(N,S,p=3)$:
\begin{equation}
N_S\left(N,S,p=3\right) = \left(2S+1\right) \min(2S+1,N/2-S+1).
\label{eq-mulst}
\end{equation}
We notice that in each $S$ sector, an ordered solution
countains a number of levels which is strictly related to the number of
sublattices of the selected order: in the lower $S$ subspace this number
is independent of the sample size for $p \le 3$ (it is the
Hilbert space dimension of a rotator or a 
symmetric top for $p=2$ (respectively $p=3$)).
In each $S$ subspace, amongst the $N_S(N,S,p=3)$ levels, the number
of appearence ($n_{\Gamma_i}^{(S)}$) of the $\Gamma_i$ IR 
can be computed following Refs. \cite{bllp94,lblp295}: 
\begin{equation}
n_{\Gamma_i}^{(S)} = \frac{1}{6} \sum_k {\rm Tr} \left(R_k|_S\right)
\chi_i\left(k\right) N_{{\rm el}}\left(k\right)
\end{equation}
where the summation index $k$ runs through the classes of 
the $S_3$ group (isomorphic to $C_{3v}$); $\chi_i\left(k\right)$,
$N_{{\rm el}}\left(k\right)$ denotes respectively the characters 
of the $\Gamma_i$ IR and the number of elements in the class $k$
(see Table \ref{table-1}). The traces of the permutations of $S_3$
in the $S$-subspace,  denoted ${\rm Tr} \left(R_k|_S\right)$,
are determined as: 
\begin{equation}
{\rm Tr} \left(R_k|_S\right) = {\rm Tr} \left(R_k\bigg|_{S_z=S}\right) - 
{\rm Tr} \left(R_k\bigg|_{S_z=S+1}\right),
\end{equation}
and
\begin{equation}
\left\{   
\begin{array}{lll}  
\displaystyle {\rm Tr} \left(I_d\bigg |_{S_z} \right) &=& 
\displaystyle   \sum_{t,v,x=-N/6}^{N/6} \delta_{t+v+x ,S_z} \\ 
\displaystyle {\rm Tr} \left((A,B)\bigg |_{S_z} \right) &=&
\displaystyle   \sum_{t,v=-N/6}^{N/6} \delta_{2t+v,S_z} \\   
\displaystyle {\rm Tr} \left((A,B,C)\bigg |_{S_z} \right) &=&
\displaystyle   \sum_{t=-N/6}^{N/6} \delta_{3t,S_z},
\end{array}
\right.
\end{equation}
where $(A,B)$ (respectively $(A,B,C)$) stands for a two-body
(respectively three-body) permutation of  $S_3$. 
The final results of the computation of the $n_{\Gamma_i}^{(S)}$
are given for the $N=27$ sample in Table \ref{table-2}.

The lowest levels in the first $S$ subspaces of the $N=27$ sample 
spectrum are given in Table \ref{table-3}. The levels which have the good 
symmetries to describe a $\sqrt 3 \times \sqrt 3$ antiferromagnet
are displayed with an asterisk: most of these $N_S (N=27, S, p=3)$ 
levels are rather far in the spectrum
and many levels belonging to other S.G.I.R. proliferate between them. 
In fact, in the odd $N$ samples, the number of low lying
$S=1/2$ states
below the first $S=3/2$ states
increases very rapidly with the system size, the trend seems the same  
in the even samples (see Fig. \ref{fig:dens}).
Altogether these numbers of  low lying levels grow seemingly 
roughly as $\alpha^N$ with $\alpha \simeq 1.18$ 
(resp. $\alpha \simeq 1.14$) in the 
odd (resp. even) samples. 
Note that these numbers lay between the  ground-state degeneracy of the 
three states Potts model \cite{ba70} ($\alpha \simeq 1.134$) and 
the degeneracy of the Dimer model \cite{e89}
($\alpha = 2^{1/3} \simeq 1.26$).   
This exponential
proliferation of  low lying 
levels with all spatial symmetries is certainly the
deepest proof of the absence of long range antiferromagnetic order: it
signs both the absence of a finite number of ordered sublattices (that
is the absence of an antiferromagnetic order parameter) and the
impossibility of a N\'eel symmetry breaking.

The selection of order by  quantum fluctuations previously observed
in the
$J_1-J_2$ model on a triangular lattice is quite different\cite{lblp295}.
In that  last model the four-sublattice QDJS family is perfectly pure on
the smallest non frustrating sample spectra:
the two-sublattice QDJS family is a subset of the
first family. 
Quantum fluctuations just stabilize this subset relatively
to the entire four-sublattice family
and thus build a simpler structure with
an order parameter of higher symmetry:
They do not create the order parameter,
but just renormalize it and increase its intrinsic symmetry.
 
\section {Ordering with a second neighbor ferromagnetic exchange
coupling}
In order to ascertain our conclusion and reinforce the credibility
of the method, we have studied with the same protocol the problem of the
kagom\'e lattice with a first neighbor antiferromagnetic 
interaction $J_1$ and a
second neighbor ferromagnetic interaction $J_2$ 
(see Eq. (\ref{eq-ham})) favouring the existence
of a $\sqrt 3\times \sqrt 3$ order. For 
large enough $J_2<0$, the spectrum has all 
the expected features of a ``Pisa Tower'' of QDJS (dynamics, number of
states and symmetries) associated with this order (see Fig. \ref{fig:j29} 
for the $N=9$ sample). When $|J_2|/J_1$
decreases the ``Pisa Tower'' disappears, indicating the existence of
a quantum phase transition.
The estimate of
the critical value of $J_2/J_1$ is a difficult task 
requiring a study of finite size effects. 
The smallest sizes compatible with the $\sqrt 3\times \sqrt 3$
order are $N=9,27,36$ sites: computing time and memory requirements
for such an extensive study remain prohibitive.
However, using appropriate twisted boundary conditions ($\pm 2\pi/3$
around the $z$-axis, see Ref. \cite{bllp94} and Appendix A), we can  use
 the  intermediate $N=12,21$ samples.
The twisted boundary conditions
break the rotational spin symmetry of
the hamiltonian and fix
the N\'eel plane perpendicular to the  twist axis. 
The helicity of N\'eel order  is thus fixed along the $z$-axis and
the free dynamics of the system   reduces to the precession of the
total spin around $z$. The effective hamiltonian reads:
\begin{equation}
{\cal H}_{\rm eff} = \frac{S_{z}^{2}}{2I_{3}},
\label{eq-hamileff1}
\end{equation}
where $I_3$ is the inertia of the top along the $z$-axis.
The degeneracy of each $S_z^2$ subspace is 2 ($\pm S_z$).
The IR 
characterizing the $\sqrt 3 \times \sqrt 3$ order is 
$\Gamma_1 : \left[{\bf k} = {\bf 0}, {\cal R}_{2\pi/3}\Psi=\Psi, 
\sigma_x\Psi=\Psi \right]$. Fig. \ref{fig:j221} shows the
tower of states associated with the $\sqrt 3 \times \sqrt 3$ 
order for the $N=21$ sample with $|J_2/J_1| = 1$.

The order parameter or the spin stiffness of
these levels scale as $N^{-1/2}$,  whereas the slope of the ``Pisa
Tower'' scales as $N^{-2}$.  At first sight, it seems  thus 
more efficient 
to use this information to look at the transition between order and
disorder as $|J_2|/J_1$ is decreased. This can be done in a rather
naive way by looking at the spin gap between the  ground-state of
$S_{z,{\rm min}}$ and $S_{z,{\rm min}}+1$: in the ordered phase
this quantity should scale as $N^{-1}$ (respectively $N^{-1/2}$ and
$N^{0}$ in the critical and disordered phase): this study is done in
Fig. \ref{fig:gap} for the case $|J_2/J_1| = 0$ (pure KAH).
It shows thanks to
the trends in the larger sample  sizes that the KAH is
certainly not ordered, probably not critical but truly disordered.

This naive use of the ``Pisa Tower'' does not account  for the whole
qualitative information contained in the QDJS family
which is described  both by the effective hamiltonian and by the space
group symmetry of the levels.
This information is incorporated in the index  $R$ measuring the
``degree of order'' present in the low lying levels, 
and defined as follows:

$i)$ First, we consider the ordered phase (with $|J_2/J_1| \geq  1$)
and concentrate on all the low lying levels of the ``Pisa Tower''
that are lower  in energy than the
softest magnons: for each sample size this determinates the number of
$S_z$ sectors which give a consistent picture of the
``quasi-classical ground-state''. Precisely we search for the lower 
level of the spectrum with a non zero wave vector in the magnetic
Brillouin zone ($E_{min}( {\bf k})$) and determine  $S_{z,max}$ as
the largest value of  $S_{z}$ such that
$ E_0(S_{z,{\rm max}}+1) \leq E_{min}( {\bf k})$.
By definition $E_0(S_{z,{\rm max}}+1)$ is thus smaller than the
energy of the softest magnon of the sample and in the thermodynamic
limit
 $S_{z,{\rm max}} $ grows roughly as $N^{1/4}$.

$ii)$ In each  $S_z$ subspace considering all the levels in the energy
range [$E_0(S_z)$,$E_0(S_z+1)$] we define $r_{S_z}$  as the ratio of
the number of  levels  compatible with $\sqrt 3\times \sqrt 3$ order
to the total number of levels of this  range ($r_{S_z}$ is 
a number between $0$ and $1$).

$iii)$ We then compute the index $R$ 
measuring the degree of order of the sample 
as the average of the ratios $r_{S_z}$ for $S_z$ running from
$0$ (resp. $1/2$) up to $S_{z,{\rm max}}$. 

This index is equal to one if the lowest levels form 
a  true ``Pisa Tower'' and  decreases 
when other IRs appear in the low lying levels 
of the spectrum when $|J_2/J_1|$ is decreased. 
 $R$ is thus a measure of the breakdown of order which 
includes qualitative information on the low lying levels. It
stands 
on the quantities that scale the more rapidly with the  system size.

The variations of 
this index as a function of $|J_2|/J_1$ and of the 
sample size 
are given in Fig. \ref{fig:index}. When $|J_2/J_1| = 0$ this index goes
very rapidly to zero  with $N$:
this conforts
 the idea that the KAH is indeed disordered and that quantum
fluctuations
show no tendency to select ``order from disorder'' in the disordered phase.
More unexpected, the finite size scaling on $R$ indicates that 
a small ferromagnetic exchange is not sufficient to
establish long range order in the system and that the 
value of the critical ratio $(|J_2|/J_1)_c$ is probably larger than $0.5$. 

\section{Incommensurate magnetic order}

\begin{figure}
\caption{
Variations of the ``spin gap'' with boundary conditions:
the small horizontal
tick gives the value of the gap for periodic boundary conditions.
There is 
a systematic effect: the $\Delta S=1$ excitation energy decreases
with the twists of the boundary conditions in the even $N$ samples and
increases in the odd $N$ ones.
}
\label{fig:twistedgap}
\end{figure}
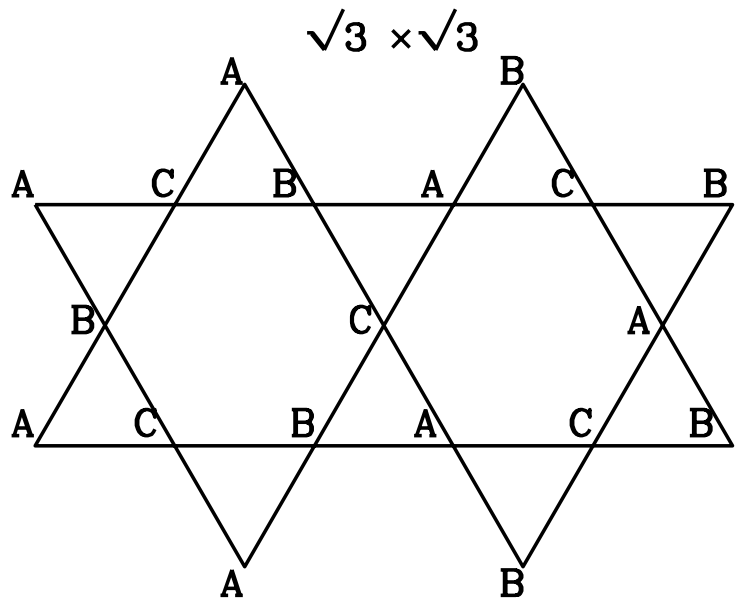
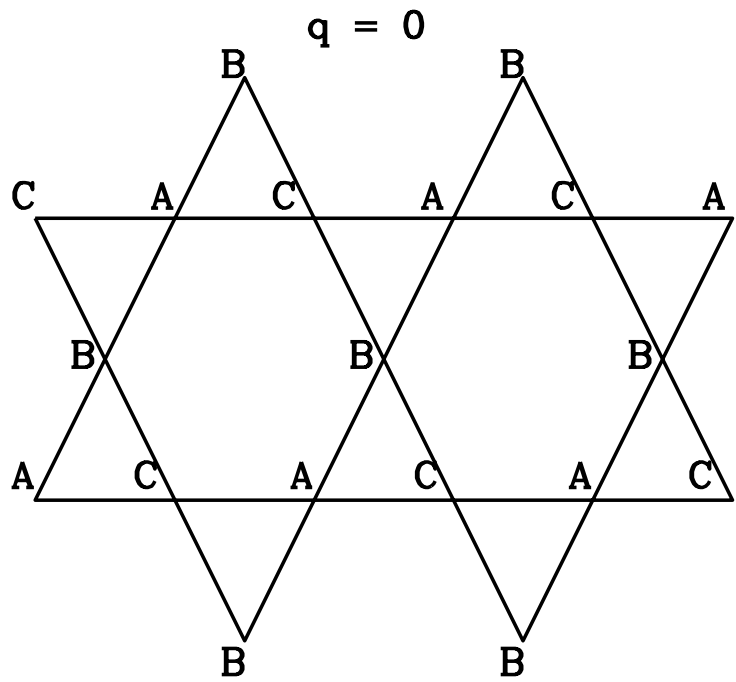
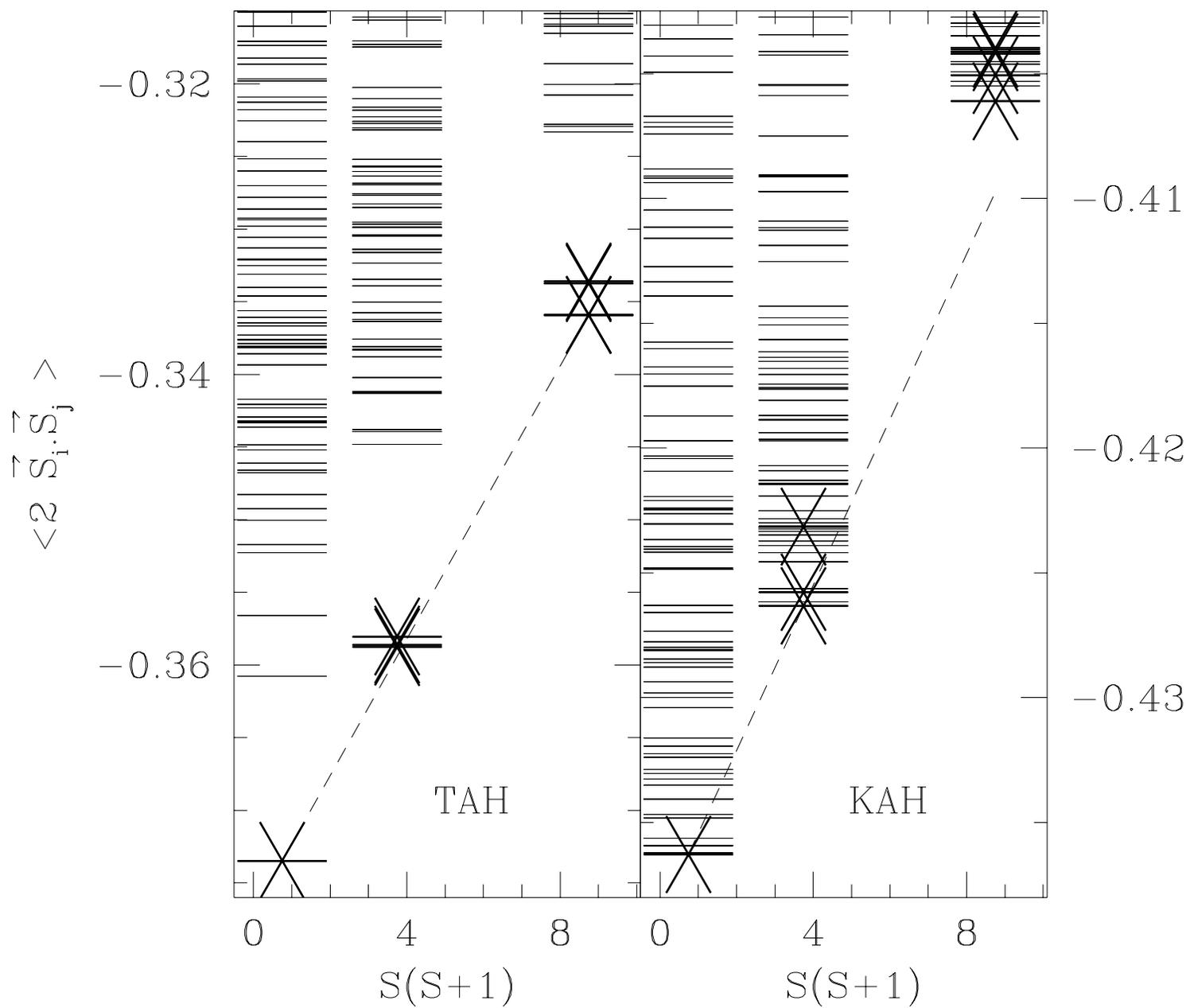
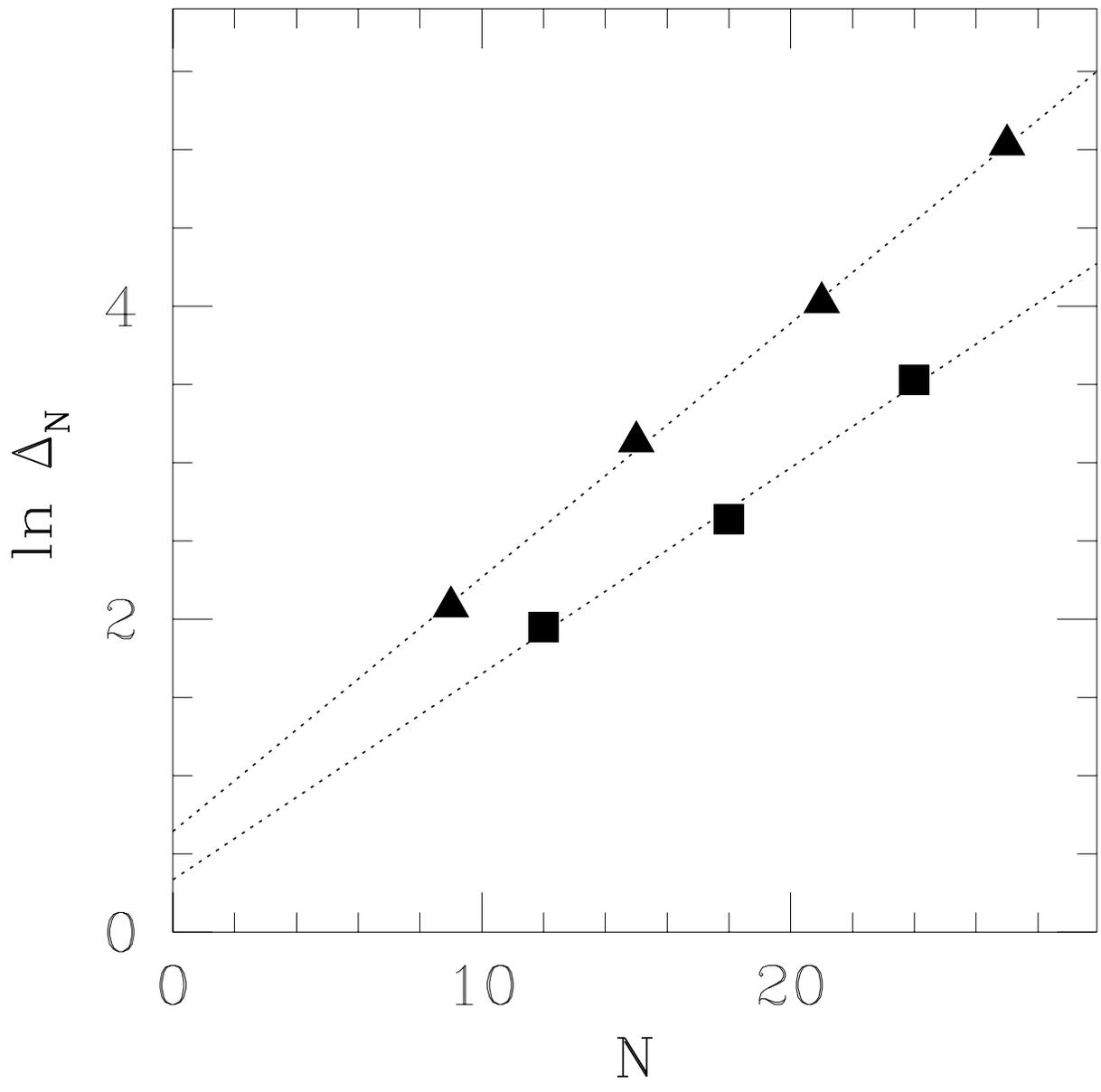
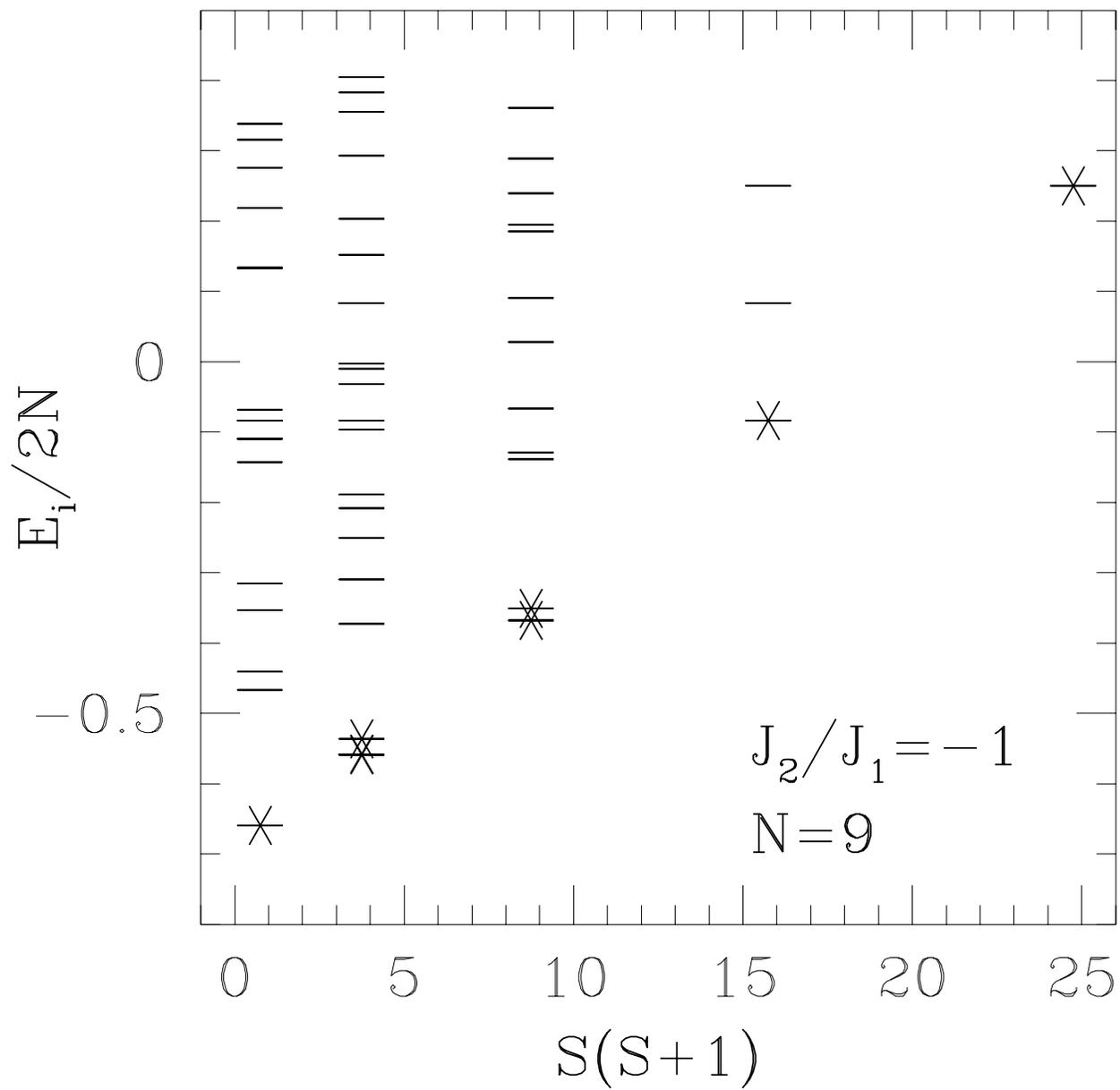
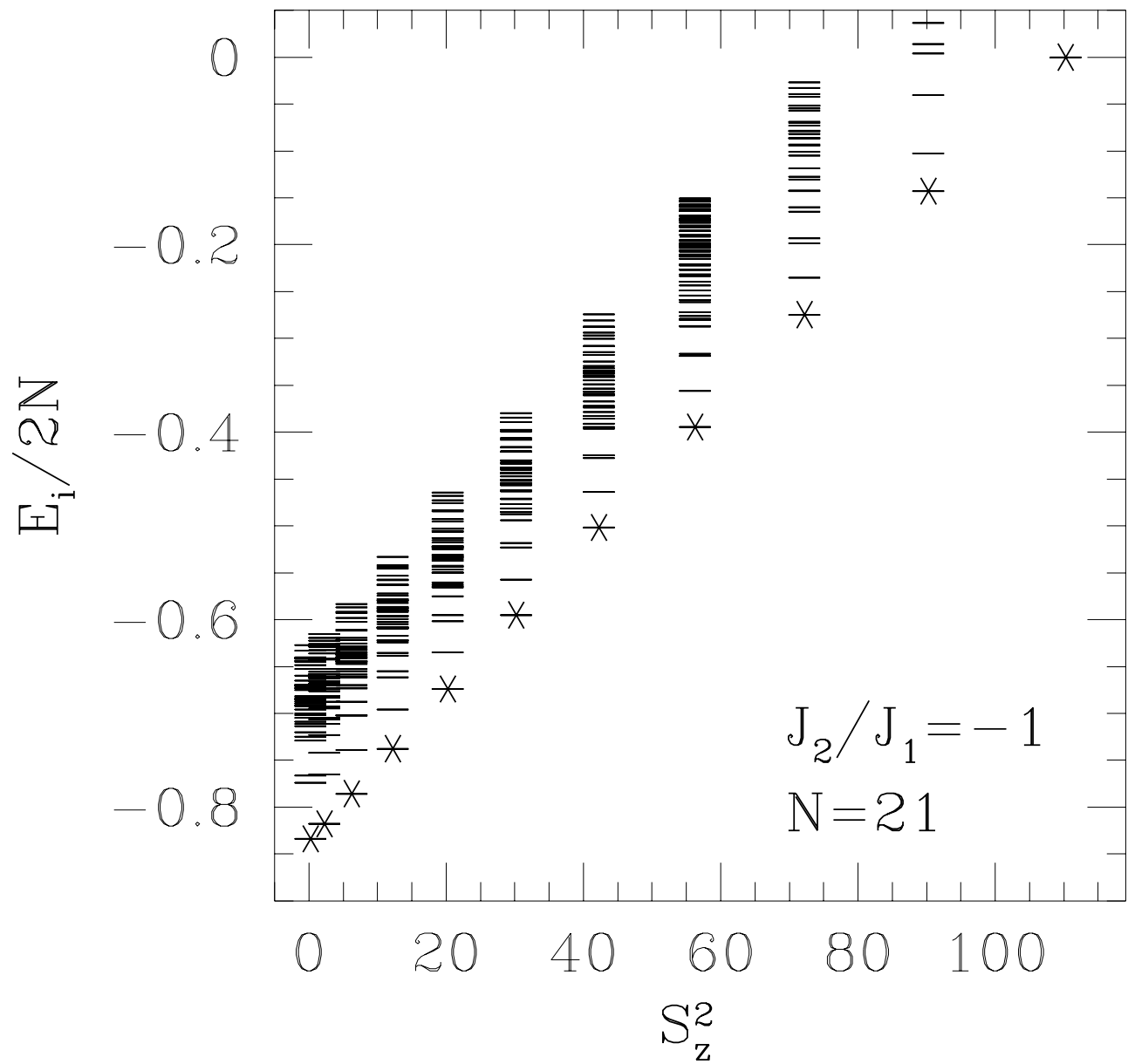
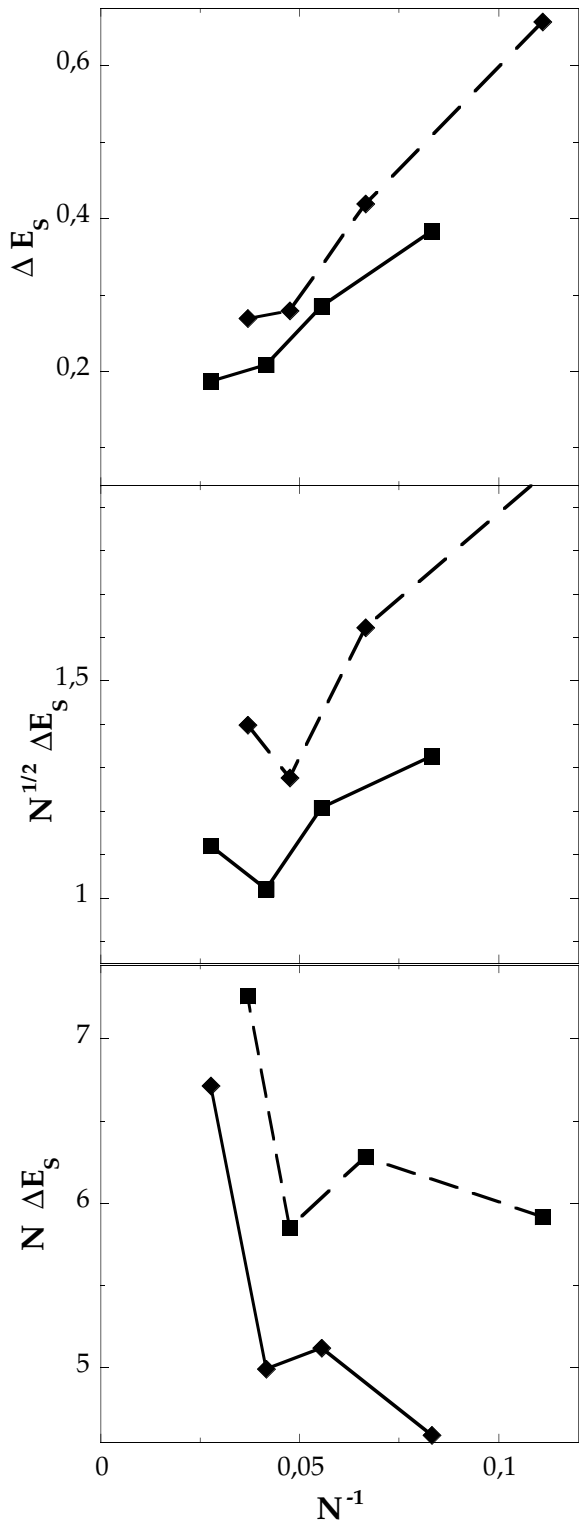
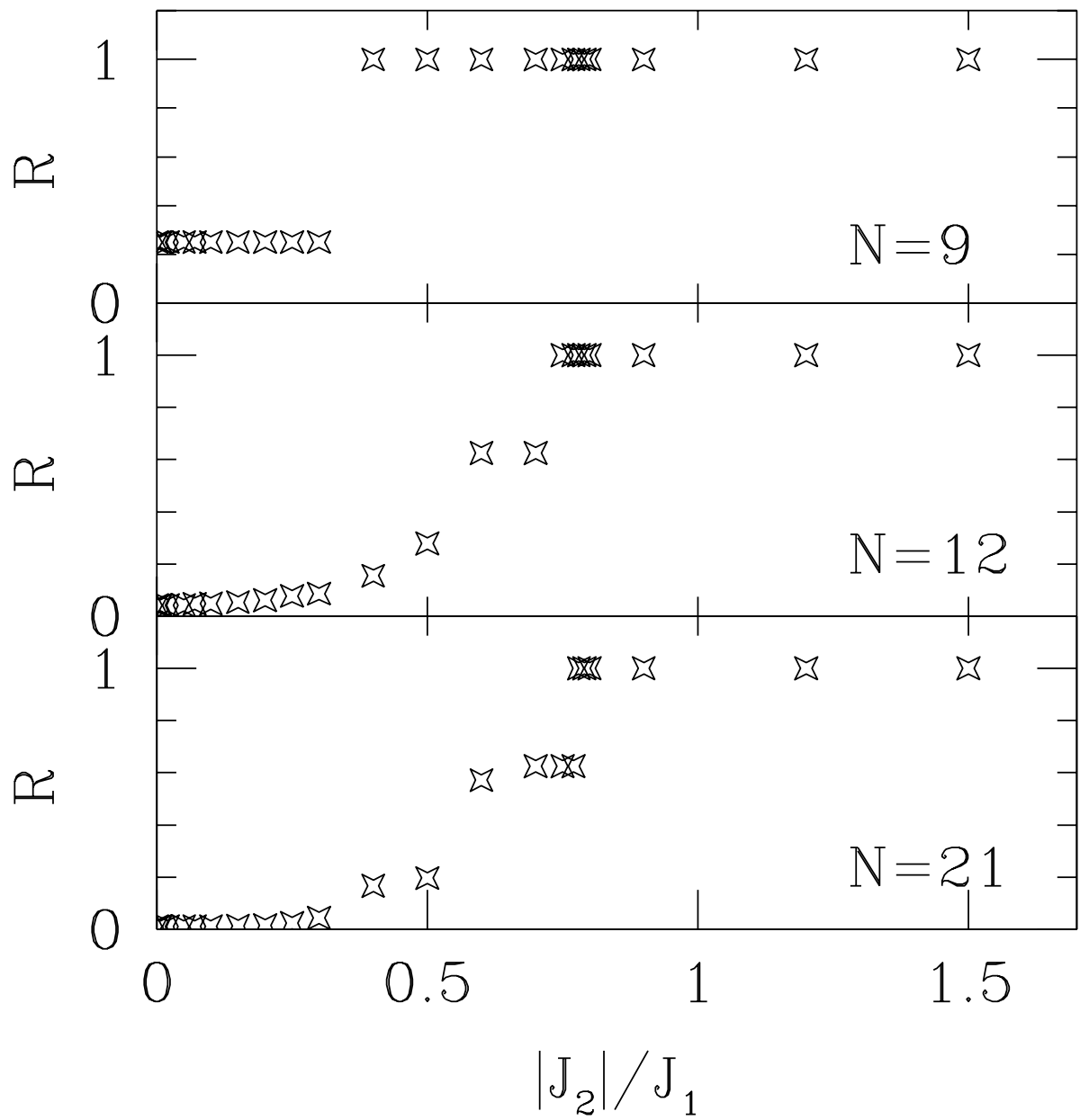
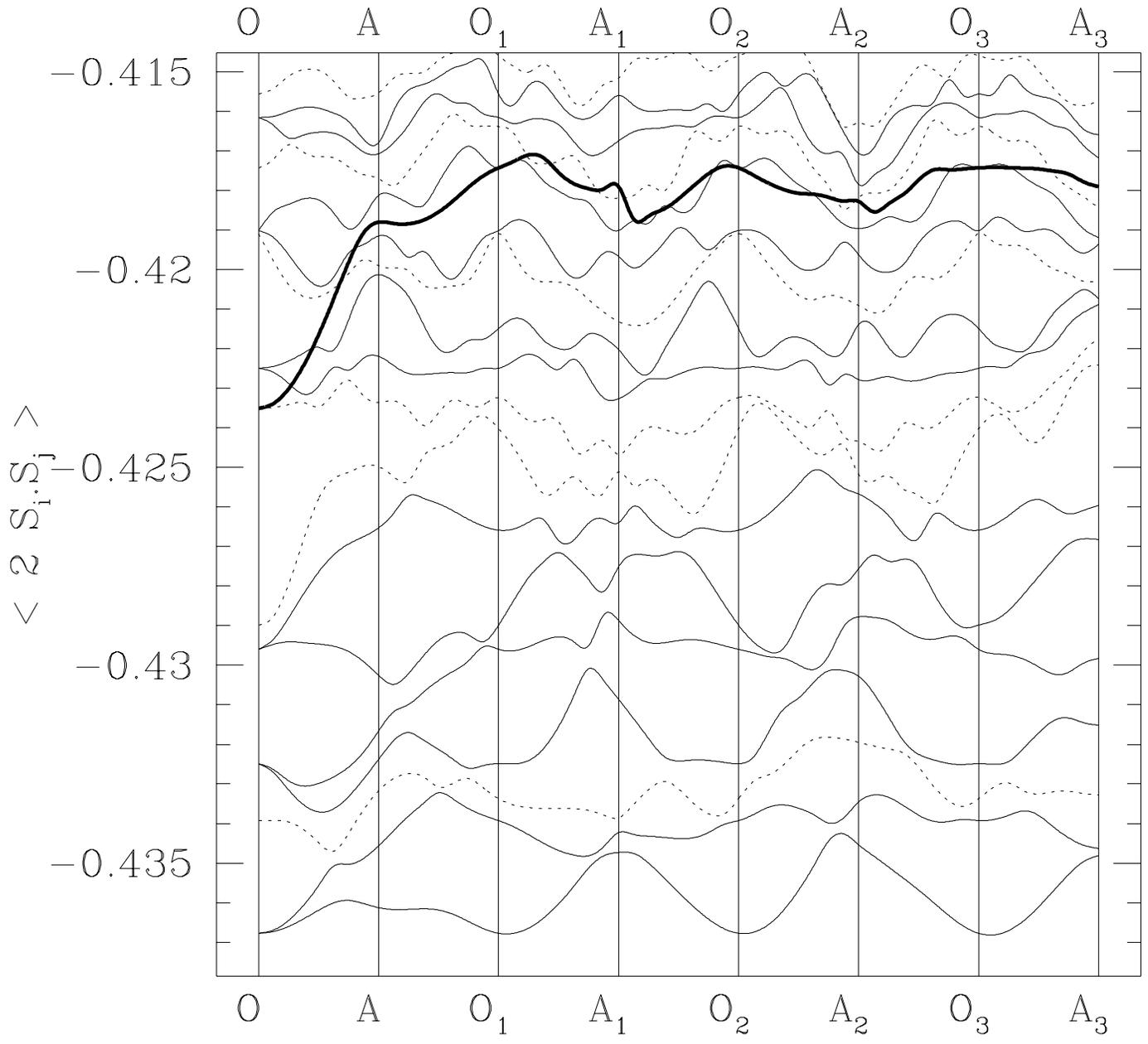
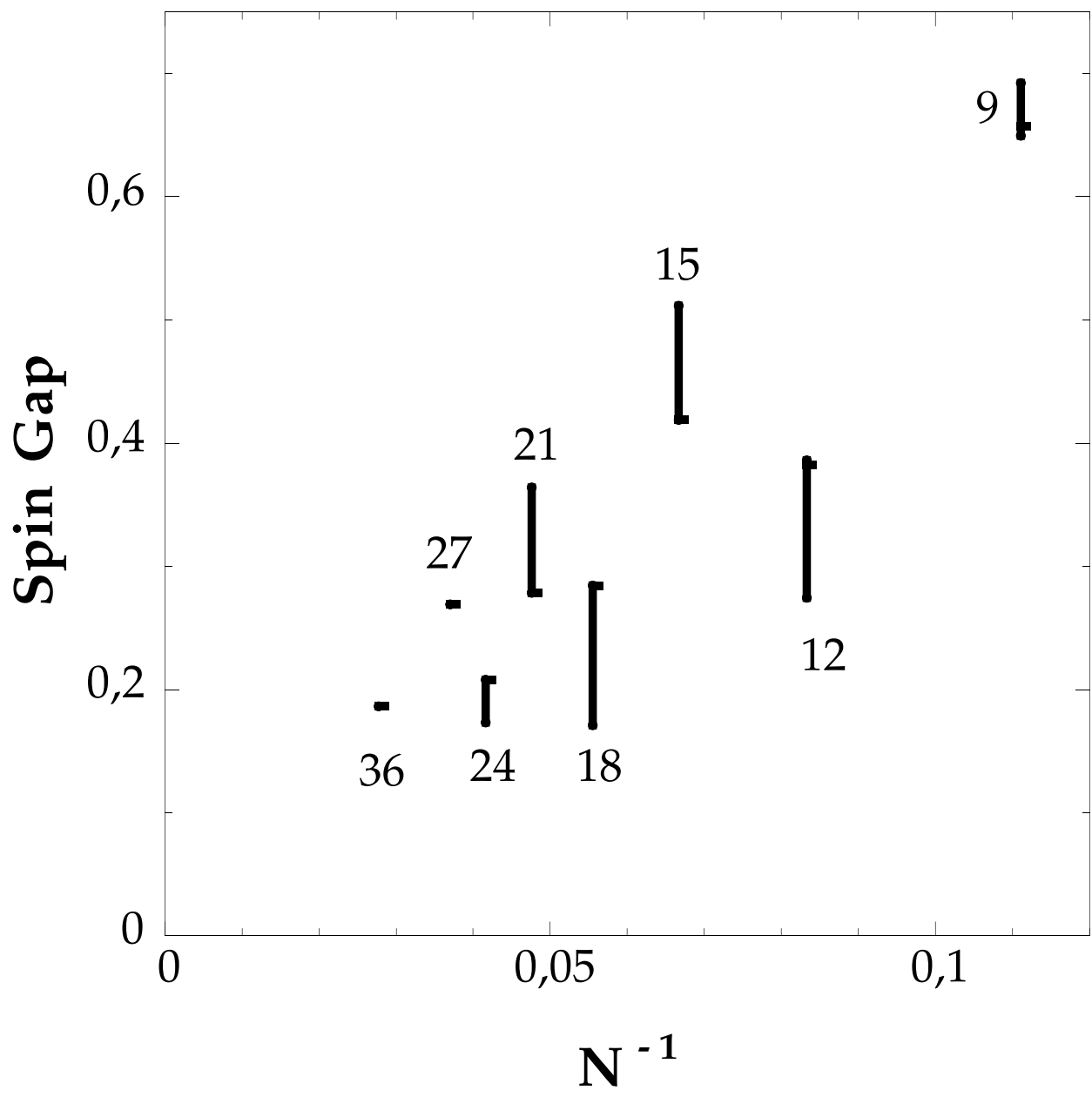

The  previous study discards 
the hypothesis of a $\sqrt 3 \times \sqrt 3$
order  in the pure KAH (we did the same check with the same conclusion
for the ${\bf q} = {\bf 0}$ order in Ref.\cite{phle95}).
Using twisted boundary conditions
${\bf S}_{{\bf  r}_i+{\bf T}_{ \alpha = 1,2}} = {\cal R}_{\bf z}(\Phi_{ \alpha = 1,2}) 
{\bf S}_{{\bf r}_i}$ across the sample defined by the vectors
${\bf T}_{ \alpha = 1,2}$ (see the appendix for more details),
we have extended our search to any  planar  antiferromagnetically  ordered
configurations, either commensurate or incommensurate. 
The existence of a planar order, if any, would be  signed by a minimum of
the   ground-state energy for a given  pair of twist angles ($\Phi_1$, $\Phi_2$)
and the appearence of a ``Pisa Tower'' for this couple of parameters \cite{bllp94}.  
A typical result of a set of diagonalizations is shown in 
Fig. \ref{fig:twist}  for the
$N=21$ sample. Sweeping the Brillouin zone for ($\Phi_1$, $\Phi_2$),
we have studied in this way the spectra of the $N=9,12,15,18,21,24,27$ samples.
We observe the following properties:

\noindent
$i)$ the influence of the twisted boundary conditions is very small,
much smaller than for the TAH:
 on the $N=21$ sample the effect of the twist on the  ground-state
of the  KAH is  only $8\%$ of the same effect on the TAH.
This is coherent with the picture of a 
disordered, liquid system,\\ 
$ii)$ we do not find a signature of any planar  antiferromagnetic 
order for any size. Very
shallow minima appear in the spectra, but they are never associated
with a
tower of QDJS and the position of these minima changes from place to
place with the sample size,\\
$iii)$ the ``spin-gap'' between the  ground-state 
energy of the $ S_{z} = 0$ (or $1/2$) subspace and the  ground-state
energy of 
the $S_{z}+1$ subspace  ($ \Delta E_{s} = E_{0}(S_{z}+1) -  E_{0}(S_{z})$) has only
small variations with the twists	
(Fig. \ref{fig:twistedgap}). These variations appear
systematic, and show a different trend in the even and odd samples:
this could be related to the fact that the odd samples only
accomodate  spin-1/2 excitations of the thermodynamic absolute
  ground-state
which is a true singlet. (This hypothesis is examined in a companion
paper \cite{bllpswe97}).

\section{conclusion}

Using the analysis of the low lying levels of the  Heisenberg
antiferromagnet on a kagom\'e lattice  we have shown 
new evidences that the
system has no planar antiferromagnetic long range order at $T=0$.  
Introducing a small second neighbor ferromagnetic 
exchange coupling does not seem to be sufficient
to establish long range order: from the experimental point 
of view this is good news as it could perhaps enlarge the number of
candidates for a spin liquid behaviour.

The theoretical study of the  tower of low lying levels (``Pisa Tower'')
  that we have developped
in this paper seems potentially useful to give an approximate location
of the transition from order to disorder even on small samples:
its advantage on other approaches stands on the finite size
scaling of the parameter we are looking at and on the inclusion
in this parameter of qualitative information on the macroscopic
ground-state.

According to our present results, this
spin-1/2 model exhibits a quantum critical point at  a
non zero $|J_2|/J_1$. It would be interesting to
investigate 
the universality class of this quantum critical point 
and  see how it may compare to the theoretical predictions  of
the non linear sigma model for canted antiferromagnets \cite{nlsig}. 

{\bf Acknowledgements}:
We have benefited from a grant of computer time at Centre de Calcul
Vectoriel pour la Recherche (CCVR), Palaiseau, France.

\appendix
\section{Twisted boundary conditions}

For a sample defined by the two vectors: 
\begin{equation}
\left\{
\begin{array}{lll}
\displaystyle {\bf T}_{1}=2(l+m){\bf u}_{1}+2m{\bf u}_{2}\\
\displaystyle {\bf T}_{2}=2l{\bf u}_{1}+2(l+m){\bf u}_{2}
\end{array}
\right.
\end{equation}
where ${\bf u}_1$ and ${\bf u}_2$ are two unit vectors of 
the kagom\'e lattice, and $m$ and $n$ 
are two integers related to the number of sites of the sample by:
$N = 3 (l^2+m^2+lm)$,
the boundary conditions are defined through: 
\begin{equation}
{\bf S}_{{\bf  r}_i+{\bf T}_{ \alpha = 1,2}} = {\cal R}_{\bf z}(\Phi_{ \alpha = 1,2}) 
{\bf S}_{{\bf r}_i}
\label{eq-tbc}
\end{equation}
where ${\bf r}_i$ denotes the sites of the kagome lattice.
In order to recover the translation invariance that seems to be broken by
these boundary conditions, the spin frame at 
the point ${\bf r}_i+{\bf u}_1$ (resp.
${\bf r}_i+{\bf u}_2$) is
rotated with respect to the spin frame at point 
${\bf r}_i$ by an angle $\theta_1$ (resp. $\theta_2$). The boundary angles
$\Phi_{ \alpha = 1,2}$ are related to $\theta_{ \alpha = 1,2}$ by the relations:
\begin{equation}
\left\{
\begin{array}{lll}
\displaystyle \Phi_{1} = 2(l+m) \theta_{1} + 2m \theta_{2}\\
\displaystyle \Phi_{2} = 2l \theta_{1} + 2(l+m) \theta_{2}.
\end{array}
\right.
\label{eq-conecb}
\end{equation}
The hamiltonian in the new frame reads:
\begin{equation}
{\cal H} = 2J_1 \sum_{\stackrel {i=1,N}{\mu=1,3}}
{\tilde {\bf S}_{{\bf r}_i}}. {\cal R}_{z}(\theta_{\mu})
{\tilde {\bf S}_{{\bf r}_i+{\bf u}_{\mu}}},
\label{eq-hamiltor}
\end{equation}
$\theta_1$ and $\theta_2$ are changed step by step so that 
$\Phi_1$ and $\Phi_2$ sweep the
appropriate fraction of the ``Brillouin zone'' of this problem .

\newpage

\begin{thebibliography}{10}

\bibitem{m91}
E. Manousakis, Rev. Mod. Phys. {\bf 63},  1  (1991). See
references therein.

\bibitem{he88}
D.A. Huse and V. Elser, Phys. Rev. Lett. {\bf 60},  2531  (1988).

\bibitem{jlg89}
T. Jolicoeur and J.C.~Le Guillou, Phys. Rev. B {\bf 40},  2727  (1989).

\bibitem{blp92}
B. Bernu, C. Lhuillier, and L. Pierre, Phys. Rev. Lett. {\bf 69},  2590
  (1992).

\bibitem{de93}
R. Deutscher and H.U. Everts, Z. Phys. B. {\bf 93},  77
  (1993).

\bibitem{esy93}
N. Eltsner, R.R.P. Singh, and A.P. Young, Phys. Rev. Lett. {\bf 71},  1629
  (1993).

\bibitem{bllp94}
B. Bernu, P. Lecheminant, C. Lhuillier, and L. Pierre, Phys. Rev. B {\bf 50},
  10048  (1994).

\bibitem{lblp195}
P. Lecheminant, B. Bernu, C. Lhuillier, and L. Pierre, Phys. Rev. B {\bf 52},
  9162  (1995).

\bibitem{bo95}
M. Boninsegni, Phys. Rev. B {\bf 52},  15304  (1995).

\bibitem{ze90}
C. Zeng and V. Elser, Phys. Rev. B {\bf 42},  8436  (1990).

\bibitem{ce92}
J.T. Chalker and J.F.G. Eastmond, Phys. Rev. B {\bf 46},  14201  (1992).

\bibitem{le93}
P.W. Leung and V. Elser, Phys. Rev. B {\bf 47},  5459  (1993).

\bibitem{sh92}
R.R.P. Singh and D.A. Huse, Phys. Rev. Lett. {\bf 68},  1766  (1992).

\bibitem{el94}
N. Eltsner and A.~P. Young, Phys. Rev. B {\bf 50},  6871  (1994).

\bibitem{s92}
S. Sachdev, Phys. Rev. B {\bf 45},  12377  (1992).

\bibitem{mz91}
J.B. Marston and C. Zeng, J. Appl. Phys. {\bf 69},  5962  (1991).

\bibitem{ze95}
C. Zeng and V. Elser, Phys. Rev. B {\bf 51},  8318  (1995).

\bibitem{hkb92}
A.B. Harris, C. Kallin, and A.J. Berlinsky, Phys. Rev. B {\bf 45},  2899
  (1992).

\bibitem{c92}
A. Chubukov, Phys. Rev. Lett. {\bf 69},  832  (1992).

\bibitem{c193}
A. Chubukov, J. Appl. Phys. {\bf 73},  5639  (1993).

\bibitem{mtgc94}
L.~O. Manuel, A.~E. Trumper, C.~J. Gazza, and H.~A. Cecatto, Phys. Rev. B {\bf
  50},  1313  (1994).

\bibitem{as95}
H. Asakawa and M. Suzuki, Int. J. Mod. Phys. B {\bf 9},  933  (1995).

\bibitem{vbcc80}
J. Villain, R. Bidaux, J.-P. Carton, and R. Conte, J. Phys. Fr. {\bf 41},  1263
   (1980).

\bibitem{s82}
E.F. Shender, Sov. Phys. J.E.T.P. {\bf 56},  178  (1982).

\bibitem{h89}
C.~L. Henley, Phys. Rev. Lett. {\bf 62},  2056  (1989).

\bibitem{jdgb90}
T. Jolicoeur, E. Dagotto, E. Gagliano, and S. Bacci, Phys. Rev. B {\bf 42},
  4800  (1990).

\bibitem{cj92a}
A. Chubukov and T. Jolicoeur, Phys. Rev. B {\bf 46},  11137  (1992).

\bibitem{k93}
S.~E. Korshunov, Phys. Rev. B {\bf 47},  6165  (1993).

\bibitem{lblp295}
P. Lecheminant, B. Bernu, C. Lhuillier, and L. Pierre, Phys. Rev. B {\bf 52},
  6647  (1995).

\bibitem{chs92}
J.T. Chalker, P.~C.~W. Holdsworth, and E.~F. Shender, Phys. Rev. Lett. {\bf
  68},  855  (1992).

\bibitem{ccr93}
P. Chandra, P. Coleman, and I. Ritchey, J. Phys. Fr. {\bf 3},  591  (1993).

\bibitem{cd94}
P. Chandra and B. Doucot, J. Phys. A {\bf 27},  1541  (1994).

\bibitem{dh92}
J. von Deft and C.~L. Henley, Phys. Rev. Lett. {\bf 69},  3236  (1992).

\bibitem{dh93}
J. von Deft and C.~L. Henley, Phys. Rev. B {\bf 48},  965  (1993).

\bibitem{f89}
D.S. Fisher, Phys. Rev. B {\bf 39},  11783  (1989).

\bibitem{nz89}
H. Neuberger and T. Ziman, Phys. Rev. B {\bf 39},  2608  (1989).

\bibitem{adm93}
P. Azaria, B. Delamotte, and D. Mouhanna, Phys. Rev. Lett. {\bf 70},  2483
  (1993).

\bibitem{a52}
P.W. Anderson, Phys. Rev. {\bf 86},  694  (1952).

\bibitem{lh95}
G. Levine and J. Howard, Phys. Rev. Lett. {\bf 75},  4142  (1995).

\bibitem{lm62}
E. Lieb and D. Mattis, J. Math. Phys. {\bf 3},  749  (1962).

\bibitem{phle95}
P. Lecheminant, PhD thesis, Universit\'e  Pierre et Marie Curie, Paris, 1995.

\bibitem{ba70}
R.~J. Baxter, J. Math. Phys. {\bf 11},  784  (1970).

\bibitem{e89}
V. Elser, Phys. Rev. Lett. {\bf 62},  2405  (1989).

\bibitem{bllpswe97}
B. Bernu, P. Lecheminant, C. Lhuillier, L. Pierre, P. Sindzingre,
C. Waldtmann, H.U. Everts, submitted to Phys. Rev. Lett. (1997).

\bibitem{nlsig}
P. Azaria, B. Delamotte, and D. Mouhanna, Phys. Rev. Lett. {\bf 68}, 1762 
(1992). A. Chubukov, T. Senthil, and S. Sachdev, Phys. Rev. Lett. {\bf 72},
2089 (1994).  A. Chubukov, S. Sachdev, and T. Senthil, Nucl. Phys. B 
{\bf 426}, 601 (1994). 
 
\end{thebibliography}

\begin{table}
\begin{center}
\begin{tabular}{|c|c c c |}
\hline
$S_3$ & $ I $ & $(A,B,C)$ & $(A,B)$ \\
$ N_{el}$ & 1 & 2 & 3 \\
  \hline
   $\Gamma_1 $ & $ 1 $ & $ 1 $ & $ 1 $ \\
   $\Gamma_2 $ & $ 1 $ & $ 1 $ & $ -1$ \\
   $\Gamma_3 $ & $ 2 $ & $ -1 $ & $ 0 $ \\
\hline
\end{tabular}
\end{center}
\caption[99]{Character table of the permutation group $S_3$.
First line indicates classes of permutations. 
The number of elements in each class is 
$N_{el}$.}
\label{table-1}
\end{table}

\begin{table}
	\begin{center}
		\begin{tabular}{|c|c c c c c c c c c c c c c c|}
			\hline
			$N=27$ &&&&&&&&& \\
                        $2S$ & 1 & 3 & 5 & 7 & 9 & 11 & 13 & 15 & 
                        17 & 19 & 21 & 23 & 25 & 27 \\
			\hline
			$n_{\Gamma_1}^{(S)}$ & 0 & 1 & 1 & 1 & 2 & 2 & 1 & 2 &
                        1 & 1 & 1 & 1 & 0 & 1 \\
			$n_{\Gamma_2}^{(S)}$ & 0 & 1 & 1 & 1 & 2 & 1 & 1 & 1 & 
                        1 & 0 & 1 & 0 & 0 & 0 \\
			$n_{\Gamma_3}^{(S)}$ & 1 & 1 & 2 & 3 & 3 & 3 & 3 & 2 & 
                        2 & 2 & 1 & 1 & 1 & 0 \\
			\hline
		\end{tabular}
	\end{center}
\caption[99]{Number of occurrences $n_{\Gamma_i}^{(S)}$ of 
each irreducible 
representation $\Gamma_i$ ($i=1,2,3$) 
with respect to the total spin $S$.}
\label{table-2}
\end{table}

\begin{table}
  \begin{tabular}{|r|rlrlrrrr|}
   $ N=27 $ & $ <2{\bf S}_i.{\bf S}_j> $ 
   & $2S$ & deg. & \multicolumn{2}{c}{$ {\bf k} $} & ${\cal R}_{2\pi/3} $ 
   & ${\cal R}_{\pi} $ & $\sigma_x $\\
   \hline
%
&        -0.43627796  &  1 &  4 &  0 &  0 & -1 &  1 &  0 \\ 
& *      -0.43627796  &  1 &  4 &  3 &  6 &  1 &  0 &  1 \\ 
&        -0.43622206  &  1 & 12 &  0 &  3 &  0 &  0 & -1 \\ 
&        -0.43593382  &  1 & 12 &  0 &  3 &  0 &  0 &  1 \\ 
&        -0.43591527  &  1 &  8 &  3 &  6 & -1 &  0 &  0 \\ 
&        -0.43563229  &  1 & 12 &  0 &  3 &  0 &  0 &  1 \\ 
  \cline{2-9}\\
 \cline{2-9}
&        -0.42632327  &  3 &  8 &  0 &  0 & -1 &  1 &  0 \\ 
& *      -0.42632327  &  3 &  8 &  3 &  6 &  1 &  0 &  1 \\ 
&        -0.42630998  &  1 &  8 &  3 &  6 & -1 &  0 &  0 \\ 
&        -0.42615931  &  3 & 16 &  3 &  6 & -1 &  0 &  0 \\ 
& *      -0.42577308  &  3 &  4 &  0 &  0 &  1 &  1 &  1 \\ 
&        -0.42562690  &  3 &  8 &  0 &  0 & -1 &  1 &  0 \\ 
&        -0.42562690  &  3 &  8 &  3 &  6 &  1 &  0 &  1 \\ 
&        -0.42485791  &  1 &  8 &  3 &  6 & -1 &  0 &  0 \\ 
&        -0.42483901  &  1 &  4 &  0 &  0 & -1 &  1 &  0 \\ 
&        -0.42483901  &  1 &  4 &  3 &  6 &  1 &  0 &  1 \\ 
&        -0.42479279  &  1 &  4 &  0 &  0 & -1 & -1 &  0 \\ 
&        -0.42479279  &  1 &  4 &  3 &  6 &  1 &  0 & -1 \\ 
&        -0.42455077  &  3 & 24 &  0 &  3 &  0 &  0 &  1 \\ 
&        -0.42419453  &  3 & 16 &  3 &  6 & -1 &  0 &  0 \\ 
&        -0.42419310  &  3 &  4 &  0 &  0 &  1 &  1 &  1 \\ 
&        -0.42417083  &  1 &  2 &  0 &  0 &  1 &  1 &  1 \\ 
&        -0.42405358  &  1 &  2 &  0 &  0 &  1 &  1 & -1 \\ 
&        -0.42405358  &  1 &  2 &  0 &  0 &  1 & -1 &  1 \\ 
&        -0.42394796  &  1 &  8 &  3 &  6 & -1 &  0 &  0 \\ 
&        -0.42391172  &  3 & 24 &  0 &  3 &  0 &  0 &  1 \\ 
&        -0.42372061  &  3 & 24 &  0 &  3 &  0 &  0 & -1 \\ 
&        -0.42366979  &  1 &  2 &  0 &  0 &  1 &  1 &  1 \\ 
&        -0.42347751  &  3 &  4 &  0 &  0 &  1 & -1 & -1 \\ 
&        -0.42333145  &  3 & 24 &  0 &  3 &  0 &  0 & -1 \\ 
&        -0.42324731  &  3 &  8 &  0 &  0 & -1 &  1 &  0 \\ 
&        -0.42324731  &  3 &  8 &  3 &  6 &  1 &  0 &  1 \\ 
&        -0.42318668  &  3 & 24 &  0 &  3 &  0 &  0 & -1 \\ 
&*       -0.42314189  &  3 &  4 &  0 &  0 &  1 & -1 &  1 \\ 
&        -0.42314189  &  3 &  4 &  0 &  0 &  1 &  1 & -1 \\ 
\end{tabular}
\caption
{Lowest eigenenergies (with degeneracy and quantum numbers) of
the $N=27$ sample spectrum.
Components of ${\bf k}$ are in units $6\pi/N$.
In the 3 last columns, 1 stands for invariant under the symmetry,
0 for no symmetry and -1 for a  non trivial phase factor under the
 above mentionned symmetry ($i.e.$ $e^{i2\pi/3}$ for the
rotation of $2\pi/3$ and -1 for the two others).
Stars indicate states possessing the symmetries 
associated with the $\protect{\sqrt 3} \times \protect{\sqrt 3}$
state. The  double horizontal bar indicates the omission of 
127 $S=1/2$ states before the first levels in the $S=3/2$ subspace. 
}
\label{table-3}
\end{table}

\newpage
\begin{figure}
\caption{Two classical planar states of the KAH:
the $\protect{\sqrt 3} \times \protect{\sqrt 3}$ 
and ${\bf q} = {\bf 0}$ states.}
\label{fig:res}
\end{figure}

\begin{figure}
\caption{
The low lying energy levels of the TAH and KAH spectrum of
the $N=27$ sample. The levels which possess the symmetry expected for an ordered
solution are denoted by a star. The  ``Pisa Tower'' of the TAH is easily
seen, well distinct from the first magnon excitations.
In the KAH on the contrary the levels candidate for the building of a tower
of states are mixed with other representations in
a continuum of excitations.
}
\label{fig:KAH}
\end{figure}

\begin{figure}
\caption{
Logarithm of the number $\Delta_{N}$ of  $S=1/2$ levels 
below the first $S=3/2$ level as a function of the sample size $N$
(black triangles).
This number which does not take into account the  two-fold 
magnetic degeneracy
is compared to the same quantity (square symbols)
for the even $N$ samples (i.e. number  of  $S=0$ levels 
below the first $S=1$ level ).
}
\label{fig:dens}
\end{figure}
 
\begin{figure}
\caption{Low lying spectrum of the $N=9$ sample
with $|J_2|/J_1 = 1$ and periodic boundary conditions,
as a function of $S(S+1)$. Note the ``Pisa Tower''  
associated with the $\protect{\sqrt 3} \times \protect{\sqrt 3}$ 
state.
}
\label{fig:j29}
\end{figure}

\begin{figure}
\caption{Low lying spectrum of the $N=21$ sample 
with $|J_2|/J_1 = 1$ 
as a function of $S_z^2$. A twist of $2\pi/3$ 
in the boundary conditions is applied to accomodate  
the $\protect{\sqrt 3} \times \protect{\sqrt 3}$ state 
with the sample size. Due to the boundary conditions, the ``Pisa Tower''
reduces to one IR for each $S_{z}$ value. }
\label{fig:j221}
\end{figure}

\begin{figure}
\caption{
Finite size study of the  ``spin gap''
$ \Delta E_{s} = E_0(S_{z,min}+1) - E_0(S_{z,min})$ 
of the pure KAH  ($ J_{2} =0$)
 as a function of the sample size.
This energy is 
$ {\cal O}(N^{\alpha})$ with $\alpha =  -1,- 1/2$ or $0$ whether
the system is ordered, critical or disordered.
Continuous lines (resp. broken lines)
are guides for the eye through the  even $N$ (resp. odd $N$)
results.  The comparison of the three quantities $ \Delta E_{s}$, 
$N^{1/2}\times  \Delta E_{s}$ and 
$N\times  \Delta E_{s}$ versus $N^{-1}$ favors the hypothesis of spin disorder.
}
\label{fig:gap}
\end{figure}

\begin{figure}
\caption{Behavior of the ``index 
of order'' ($R$) as a function of $|J_2|/J_1$ and of the system size.
}
\label{fig:index}
\end{figure}

\begin{figure}
\caption{
Variation of the  energy per link
$\left<2{\bf S}_{i}.{\bf S}_{j}\right>$
of the low lying
levels of the $N=21$ sample versus twisted boundary conditions $\Phi_1$
($\Phi_2$=0).
$O_{i}$ are the points where $\Phi_1$= $0\pmod{2 \pi}$.
$O_{7}$ is the first point where the twist per link is $0\pmod{2 \pi}$.
The points $A_{k}$ ($\Phi_1$=$\pi\pmod{2 \pi}$) are in the middle of
$O_{i}$ and $O_{i+1}$.
Because the figure is symmetric with respect to $A_3$, the part $A_3-O_7$ is
not represented here.
Note that
most of the levels do not come back to their original assignation after a
$ 2 \pi $ twist of the boundary conditions. In fact on this small size sample,
due to an extra-symmetry when $\Phi_i$=0,
the uniform  $\bf{k}$=$\bf{0}$   ground-state is
degenerate with the first star of $\bf{k} \neq \bf{0}$ eigenstates.
Only the $\bf{k}$=$\bf{0}$ states and their continuation  have been
shown in $OA_{1}$. The 6 other $\bf{k} \neq \bf{0}$ in $OA$ are found
by folding the $O_{i}A_{i}$ onto  $OA$.
Full lines stand for   states going continuously to
$\bf{k}$=$\bf{0}$  chiral states (complex IRs of $C_{3}$),
dashed lines stand for states going continuously to
the first star of $\bf{k} \neq \bf{0}$ eigenstates
the $\bf{k}$=$\bf{0}$   non chiral states (trivial IR of $C_{3}$);
bold line:
the first $S_{z}$=3/2 level.
}
\label{fig:twist}
\end{figure}

\end{document}